\documentclass[conference]{IEEEtran}
\IEEEoverridecommandlockouts
\usepackage{algorithm} 
\usepackage[noend]{algpseudocode}
\usepackage{cite}
\usepackage{amsmath,amssymb,amsfonts}
\usepackage{subcaption}
\usepackage{graphicx}
\usepackage{textcomp, textcase}
\usepackage{xcolor}
\usepackage{comment}
\usepackage{enumitem}
\def\BibTeX{{\rm B\kern-.05em{\sc i\kern-.025em b}\kern-.08em
    T\kern-.1667em\lower.7ex\hbox{E}\kern-.125emX}}
\begin{document}
\captionsetup{font=small,labelfont={bf,sf}}

%\title{DAGGen-ROS2: \textbf{DAG} \textbf{Gen}eration tool for \\\textbf{ROS2} applications enabling timing analysis}
\vspace{-10mm}
\title{Trace-enabled Timing Model Synthesis\\ for ROS2-based Autonomous Applications}
\vspace{-8mm}
\author{
\IEEEauthorblockN{Hazem Abaza\IEEEauthorrefmark{1}\IEEEauthorrefmark{2}, Debayan Roy\IEEEauthorrefmark{1}, Shiqing Fan\IEEEauthorrefmark{3}, Selma Saidi\IEEEauthorrefmark{2} and Antonios Motakis\IEEEauthorrefmark{1}}
\IEEEauthorblockA{\IEEEauthorrefmark{2}\textit{Technische Universität Dortmund}, \IEEEauthorrefmark{1}\textit{Huawei Dresden Research Center},  \IEEEauthorrefmark{3}\textit{Huawei Munich Research Center}}
\IEEEauthorblockA{\{hazem.abaza, selma.saidi\}@tu-dortmund.de, \{debayan.roy6, shiqing.fan, antonios.motakis\}@huawei.com}
}

\begin{comment}

\author{\IEEEauthorblockN{1\textsuperscript{st} Given Name Surname}
\IEEEauthorblockA{\textit{dept. name of organization (of Aff.)} \\
\textit{name of organization (of Aff.)}\\
City, Country \\
email address or ORCID}
\and
\IEEEauthorblockN{2\textsuperscript{nd} Given Name Surname}
\IEEEauthorblockA{\textit{dept. name of organization (of Aff.)} \\
\textit{name of organization (of Aff.)}\\
City, Country \\
email address or ORCID}
\and
\IEEEauthorblockN{3\textsuperscript{rd} Given Name Surname}
\IEEEauthorblockA{\textit{dept. name of organization (of Aff.)} \\
\textit{name of organization (of Aff.)}\\
City, Country \\
email address or ORCID}
\and
\IEEEauthorblockN{4\textsuperscript{th} Given Name Surname}
\IEEEauthorblockA{\textit{dept. name of organization (of Aff.)} \\
\textit{name of organization (of Aff.)}\\
City, Country \\
email address or ORCID}
\and
\IEEEauthorblockN{5\textsuperscript{th} Given Name Surname}
\IEEEauthorblockA{\textit{dept. name of organization (of Aff.)} \\
\textit{name of organization (of Aff.)}\\
City, Country \\
email address or ORCID}
\and
\IEEEauthorblockN{6\textsuperscript{th} Given Name Surname}
\IEEEauthorblockA{\textit{dept. name of organization (of Aff.)} \\
\textit{name of organization (of Aff.)}\\
City, Country \\
email address or ORCID}
}
\end{comment}

\maketitle
\vspace{-4mm}
\begin{abstract}
Autonomous applications are typically developed over Robot Operating System 2.0 (ROS2)	even in time-critical systems like automotive. 
Recent years have seen increased interest in developing model-based timing analysis and schedule optimization approaches for ROS2-based applications. 
To complement these approaches, we propose a tracing and measurement framework to \emph{obtain timing models} of ROS2-based applications.  
It offers a tracer based on \emph{extended Berkeley Packet Filter} that \emph{probes} different functions in ROS2 middleware and reads their arguments or return values to reason about the data flow in applications.
It \emph{combines} event traces from ROS2 and the operating system to generate a \emph{directed acyclic graph} showing ROS2 callbacks, precedence relations between them, and their timing attributes.
While being compatible with existing analyses, we also show how to model (i)~message synchronization, e.g., in sensor fusion, and (ii)~service requests from multiple clients, e.g., in motion planning. 
Considering that, in real-world scenarios, the application code might be \emph{confidential} and formal models are unavailable, our framework still enables the application of existing analysis and optimization techniques.
%
%
%However, the timing model is a prerequisite for the analysis and, hence, the important question is: \emph{how to obtain such timing models for ROS2 applications?}
%Until now, the purpose for tracing ROS2 applications has been limited to measuring computation and communication latency, which is not sufficient for timing analysis. %and debugging.
%This paper proposes \emph{DAGGen-ROS2}, a timing profiler for ROS2 applications based on \emph{extended Berkeley Packet Filter (eBPF)}. It can construct a \emph{directed acyclic graph} (DAG) showing ROS2 callbacks (synonymous to tasks) annotated with their timing attributes and the precedence relations between them, which is often the input to a timing analysis tool.
%It \emph{combines} event traces collected from the operating system and ROS2 middleware, respectively.
%The results can also be used for timing \emph{debugging}, e.g., when a callback's response time is violating the requirement, it can be reasoned if it is due to a longer waiting time or its execution time is too long.
We demonstrate our framework's capabilities by synthesizing the timing model of a real-world benchmark implementing \emph{LIDAR-based localization} in Autoware's \emph{Autonomous Valet Parking}. 
\end{abstract}

\begin{comment}
\begin{IEEEkeywords}
component, formatting, style, styling, insert
\end{IEEEkeywords}
\end{comment}

\section{Introduction}\label{sec:intro}
Developing industry-strength autonomous applications requires teams of engineers with different backgrounds.  
Robotic Operating System version~2 (ROS2) is a powerful middleware over which \emph{modular} software components can be developed and composed easily to create autonomous applications.
To leverage these benefits and the vast amount of \emph{open-source} contributions to autonomous applications, ROS2 is widely used even in \emph{time-critical} systems such as self-driving cars.

In recent years, there have been efforts to develop timing analysis and optimization approaches for ROS2-based applications, e.g., \cite{casini2019response,Tang2020RTSS,choi2021RTAS,Arafat2022DAC,Blass2021RTSS}.
These approaches typically assume that the application \emph{models} are well-defined, i.e., the execution times of and the precedence relations between ROS2 \emph{callbacks} are known.
%In many industry scenarios, application development process follows separation of concerns and only during system integration, the applications are composed.
However, in many industry scenarios, such models are not provided by application developers.
Further, during system integration, it is challenging to obtain many details---especially at the level of callbacks---due to \emph{confidentiality} reasons.

In parallel to the model-based techniques, \emph{tracing} ROS2-based applications have also gained interest.
In this context, \emph{ros2\_tracing} provides a framework based on Linux Trace Toolkit: next generation (LTTng)~\cite{bedard2022ros2tracing}. 
It has tracepoints in ROS2 functions to identify callbacks and topics and also track them during runtime.
\emph{Autoware\_Perf}~\cite{Li2022JSA} and \emph{CARET}~\cite{CARET} add more tracepoints and use trace data to measure the response time of a callback, the communication latency between a pair of callbacks, and the end-to-end latency of a callback-chain.
 
\begin{figure}[tbp]
	\centering
	\includegraphics[width=\columnwidth]{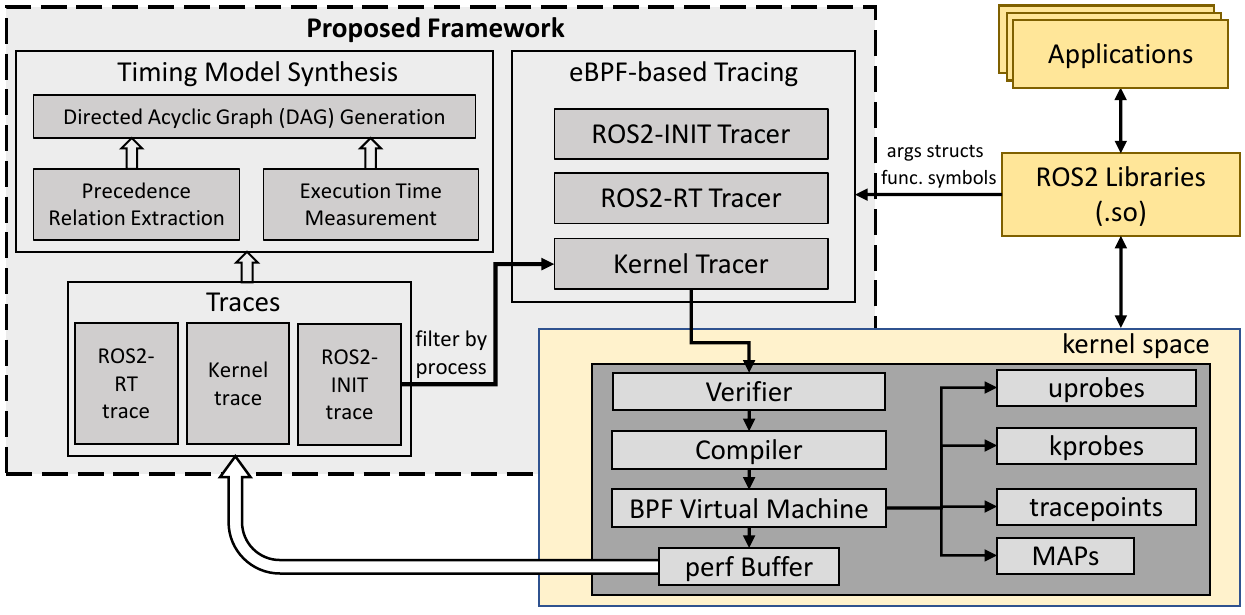}
	\vspace{-4mm}
	\caption {Proposed trace-enabled timing model synthesis framework.}
	\vspace{-7mm}
	\label{fig:framework}
\end{figure} 
 
\smallskip
\noindent\textbf{Proposed timing model synthesis framework:} This paper \emph{bridges} the gap between tracing and timing analysis.
We use \emph{extended Berkeley Packet Filter} (eBPF)~\cite{Rice2023} for tracing.
Compared to LTTng, it does not require direct instrumentation and recompilation of standard ROS2 libraries and offers efficient trace filtering. %, which is crucial for its adoption in the industry.
Unlike LTTng, eBPF does not give access to local variables in functions and, hence, we could not re-use many tracepoints identified by \cite{bedard2022ros2tracing,Li2022JSA,CARET}.
Besides probing new ROS2 functions, we traverse complex data structures of their arguments to get desired information, e.g., topic names, callback IDs, and source timestamps of data. 
Our proposed framework, in Fig.~\ref{fig:framework}, provides three tracers: 
(i)~\emph{ROS2-INIT} tracer logs the initialization of ROS2 \emph{nodes}. 
(ii)~During runtime, \emph{ROS2-RT} tracer tracks the start and end of callbacks and data read from and write to \emph{Data Distribution Service} (DDS) topics. 
Unlike \cite{bedard2022ros2tracing,Li2022JSA,CARET}, it also tracks client callbacks and message synchronization, which are found, e.g., in AUTOWARE's Autonomous Valet Parking (AVP)~\cite{AVP}.   
(iii)~\emph{Kernel} tracer logs scheduler events in the operating system (OS) related to ROS2 callbacks only with the help of eBPF's trace filtering.

As illustrated in Fig.~\ref{fig:framework}, our proposed framework uses the collected traces to synthesize timing models of applications as \emph{directed acyclic graphs} (DAGs). 
We model ROS2 callbacks as tasks (or vertices) and DDS communication between them using precedence relations (or edges).
Further, we propose to model a \emph{service} using $n$ tasks if it is invoked by $n$ different clients.
Otherwise, DAG will show one vertex with $n$ incoming edges and $n$ outgoing edges, i.e., $n \times n$ chains passing through the vertex, which is a wrong interpretation. 
Also, we propose to model a synchronization of $m$ data, e.g., for sensor fusion, using $m+1$ tasks, i.e., $m$ tasks reading data each and then these $n$ tasks feed data to another task that outputs the result.     

Further, we \emph{combine} ROS2 and scheduler events to measure the execution time of a callback for each invocation.
We get statistical information from these measurements, e.g., measured worst-case, best-case, and average values. 
We annotate the DAG with the obtained timing information. 

The DAG we generate can serve as an input for analysis and optimization by, e.g., \cite{casini2019response,Tang2020RTSS,choi2021RTAS,Arafat2022DAC,Blass2021RTSS}.
Also, our framework is \emph{not tied} to any particular application because we probe functions in ROS2 middleware and not in applications directly.
We employ our framework to synthesize the timing model of a \emph{real-world benchmark} implementing LIDAR-based localization in AVP. 
\section{Preliminaries}\label{sec:background}
\begin{comment}
\begin{itemize}
	\item What are the different elements of ROS2 applications? e.g., node, callbacks (timer, subscriber, services and actions)
	\item How ROS2 nodes communicate over DDS? as we are focusing on tracing, we should mention the different levels through which data passes in ROS2, e.g., RCL, RMW, and DDS; later we need to justify why we put an event in certain functions
	\item ROS2 computation chains: say how we define a computation chain built with ROS2 callbacks (different types of callbacks can make a chain)
	\item ros2\_tracing: mention what is possible with ros2\_tracing, mention some of the events that have been used even in out work
	\item Give definition of different timing properties, e.g., response time of a task (task = callback), execution time, period of timer callbacks, end-to-end latency and jitter of computation chains  
\end{itemize}
\end{comment}

%\noindent
%\textbf{ROS2 architecture and semantics}: 
\subsection{ROS2 architecture and semantics}\label{sec:semantics}
We study autonomous applications developed over ROS2.
They comprise ROS2 \emph{nodes} where each performs a particular function, e.g., localization and object detection.
Nodes communicate using \emph{topics}. 
%After performing certain computations, 
A node publishes data on a topic, e.g., a ``point cloud filter'' node can publish ``filtered point cloud'' data on a topic.
A node can subscribe to a topic and when there is new data on that topic, it will trigger a function, called \emph{subscriber callback}, to handle the data.
For example, the ``point cloud filter'' node subscribes to LIDAR data, and the filtering function is triggered by a new data coming from the LIDAR.
Further, a node can run a \emph{timer callback} that is triggered by a periodic timer signal.
ROS2 also offers blocking and non-blocking remote procedural calls (RPC) via \emph{services} and \emph{actions}, respectively. 
Here, a function in a server node can be directly invoked from a client node.
In this paper, we do not study actions.
Services are implemented using topics, i.e., an RPC is made by publishing data on a \emph{request} topic, and the results of the RPC are written on a \emph{response} topic.

Overall, in software developed following ROS2 semantics, we can find timer, subscriber, service, and client callbacks.
We assume that a \emph{single-threaded executor} dispatches all callbacks in a node. 
Hence, a thread executes one callback at a time from start to end before looking again into the queue with ready-to-run callbacks~\cite{casini2019response}.
For each callback $cb_k$, we measure its \emph{execution time} $et_{k,u}$ at each invocation $u \in \mathbb{N}$. 
In Sec.~\ref{sec:modelsyn}, we provide a technique to measure it.
%Based on these measurements, we can provide estimates of worst-case, best-case, and average execution times.%, which are important for timing analysis and optimization.

Data communication via a topic occurs through multiple abstraction layers in ROS2. 
For this work, we use ROS2 Foxy.
Officially-supported client libraries like \emph{rclcpp} and \emph{rclpy} comprise APIs (application programming interfaces) to develop ROS2 applications in C++ and python, respectively. 
These libraries use the core C interfaces provided by the \emph{rcl} library which implements ROS2 concepts.  
All communications between ROS2 nodes are carried out using DDS mechanisms. 
We have used Eclipse Cyclone DDS by selecting the appropriate \emph{rmw} (ROS MiddleWare) interface library.
%Different DDS implementations including eProsima FastRTPS, Eclipse Cyclone DDS, or RTI Connext DDS can be used beneath the rcl library by just selecting an appropriate \emph{rmw} (ROS MiddleWare) interface library. 
DDS layer is directly over Linux version 5.4 in our implementation. %the OS, i.e., Linux  (e.g., Linux, Windows, or macOS). 
ROS2 layered architecture is illustrated in \cite{bedard2022ros2tracing,aws}, while \cite{Kronauer2021LatencyAO} shows the communication between two nodes through these layers.

\subsection{extended Berkeley Packet Filter (eBPF)}
%
	%\item what is the principle of eBPF?
eBPF technology enables to run validated programs in kernel space without the need to recompile the kernel.
\emph{BPF Compiler Collection} (BCC) tool allows writing eBPF programs in restricted C that are converted into bytecode using Low-Level Virtual Machine (LLVM) Clang compiler.
The bytecode is loaded into the kernel using \emph{bpf()} syscall.
eBPF programs can be attached to the user- and kernel-space functions using uprobes and kprobes, respectively, thereby enhancing system observability.
In this work, we have used BCC version 0.26.0 and LLVM-clang version 10.0.	
	
%eBPF is a Linux kernel technology facilitating the execution of in-kernel virtual machine programs (eBPF programs) within the kernel context ~\cite{}. eBPF programs are loaded as bytecode into the kernel by leveraging the bpf() system call.
%However, before being loaded eBPF programs undergo a validation process through the kernel static analyzer to assesses their safety and adherence to predefined criteria, ensuring their integrity and security for the kernel environment. 
%In practice, tool like bcc is used to write efficient ebpf programs in restricted C which are then converted to bytecode using LLVM Clang compiler~\cite{}.
%eBPF programs can be hooked when various kernel and userspace functions are invoked, thereby enabling comprehensive system monitoring.
%Through the utilization of the kernel probes (kprobes) and the userspace probes (uprobes), eBPF programs can capture information regarding system calls, function entry and exit points, and network events. 
%This collected data can not only be used for monitoring but can also be efficiently shared with userspace applications through the map’s infrastructure of the eBPF.
	%\item why we choose eBPF compared to other tracers?

The primary reasons to use eBPF instead of LTTng (as in \emph{ros2\_tracing}~\cite{bedard2022ros2tracing}) are as follows:
(i)~Using eBPF, we can attach probes to the entry and exit of ROS2 functions and read their arguments and return values, respectively, without modifying the standard libraries.
%That is, we do not modifying the standard libraries.
Unlike with LTTng, there is no need for direct instrumentation and recompilation of ROS2 libraries. % as is typically the case with LTTng userspace tracing. 
\cite{CARET} has proposed to use \emph{LD\_PRELOAD} to redirect function calls to its tracing library instead of ROS2 libraries and then calls original ROS2 functions from there.
However, this requires running several lines of code to update addresses to find the original functions, which adds significant tracing overheads without any additional capabilities.
Also, such a redirection is not possible when a function is defined in the header.
(ii)~Trace filtering is more efficient in eBPF~\cite{sharma2016enhanced}, which we use to obtain kernel events specific to ROS2 nodes only. 
Further, it is possible to filter events pertaining to one or more ROS2 nodes, which is useful for quick debugging. 
(iii)~eBPF is safer and offers more programmability~\cite{Rice2023} that can be explored in the future, e.g., dynamic scheduling of ROS2 nodes to improve timing performance and secure execution of ROS2 callbacks. 

\vspace{-3mm}
\section{Proposed Trace Points}\label{sec:tracer}
\subsection{Trace points in ROS2} %We have used eBPF for tracing. 
Using eBPF, we have attached \emph{uprobes} and \emph{uretprobes} to several functions in different ROS2 layers as shown in Table~\ref{table:ROS2probes}. 
%Using a probe, we not only know when the probed function is entered (or exited) but also we can read function arguments (or return values). 
%Considering that a middleware function is called by all ROS2 nodes (or callbacks), we distinguish between them using the arguments passed on to the function.
We distinguish between callbacks (or topics) during a ROS2 function call using the arguments passed on to the function.
Here, the main challenge is to understand complex ROS2 data structures and identify functions to probe so that we get the required events with enough information to reason about what is happening at the application layer. 
%All information we extract by parsing the function arguments and return values is provided in Table~\ref{table:ROS2probes}.
Table~\ref{table:ROS2probes} lists the information we extract using each probe.
Due to limited space, we do not elaborate on how we traverse through the function arguments. 
%We highlight that we do not want to attach probes to applications because we want to develop a framework that can generically help in understanding the timing behavior of any set of applications implemented over ROS2.
%
\begin{table}[t!]
	\renewcommand*{\arraystretch}{1.1}
	\centering
	\caption{Inserted probes in ROS2 Foxy.}
	\vspace{-2mm}
	\begin{tabular}{|c|p{1.1cm}|p{1.9cm}|p{3.5cm}|} 
		\hline
		\textbf{No.} & \textbf{ROS2 lib} & \textbf{Function} & \textbf{Params}/\textbf{Purpose} \\
		\hline\hline
		$\mathbb{P}_1$ & rmw\_ cyclone dds\_cpp & rmw\_create\_ node & shows node name and the PID of the thread that will execute node's callbacks\\ \hline
		$\mathbb{P}_2$ & rclcpp & execute\_timer & notifies timer CB starts\\ \hline
		$\mathbb{P}_3$ & rcl & rcl\_timer\_call & shows timer CB ID\\ \hline
		$\mathbb{P}_4$ & rclcpp & execute\_timer & notifies timer CB ends\\ \hline
		$\mathbb{P}_5$ & rclcpp & execute\_ subscription & notifies subscriber CB starts\\ \hline
		$\mathbb{P}_6$ & rmw\_ cyclone dds\_cpp & rmw\_take\_int & notifies a read event on a topic and shows subscriber CB ID, topic name and source timestamp of data\\ \hline
		$\mathbb{P}_7$ & message\_ filters & operator & shows that a subscriber CB is used for data synchronization\\ \hline
		$\mathbb{P}_8$ & rclcpp & execute\_ subscription & notifies subscriber CB ends\\ \hline
		$\mathbb{P}_9$ & rclcpp & execute\_service & notifies service CB starts\\ \hline
		$\mathbb{P}_{10}$ & rmw\_ cyclone dds\_cpp & rmw\_take\_ \hspace{0.3cm} request & notifies a service request received event and shows service CB ID, service name, and source timestamp of request\\ \hline
		$\mathbb{P}_{11}$ & rclcpp & execute\_service & notifies service CB ends\\ \hline
		$\mathbb{P}_{12}$ & rclcpp & execute\_client & notifies client CB starts\\ \hline
		$\mathbb{P}_{13}$ & rmw\_ cyclone dds\_cpp & rmw\_take\_ \hspace{0.3cm} response & notifies a service response received event and shows client CB ID, service name, and source timestamp of response\\ \hline
		$\mathbb{P}_{14}$ & rclcpp & take\_type\_ erased\_response & notifies if a client CB will be dispatched\\ \hline
		$\mathbb{P}_{15}$ & rclcpp & execute\_client & notifies client CB ends\\ \hline
		$\mathbb{P}_{16}$ & cyclone dds & dds\_write\_impl & notifies a write event on a topic and shows the topic name and source timestamp of data/request/response \\ \hline
	\end{tabular}
	\vspace{-6mm}
	\label{table:ROS2probes}
\end{table}
We list below several important aspects of our tracer and concepts we have used while adding tracepoints.
\begin{itemize}[leftmargin=*]
	\item Using Linux \emph{perf} tool, we obtain call graphs for ROS2 nodes while running different types of callbacks. This helps to identify ROS2 functions that are called and can be probed.
	\item Each event generated by our probes comprises (i)~a timestamp for chronological ordering, (ii)~a process ID (PID) to associate the event to a ROS2 node, and (iii)~a probe name to indicate the type of information we can get from the event. 
	\item We attach \emph{uprobes} $\{\mathbb{P}_2, \mathbb{P}_5, \mathbb{P}_9, \mathbb{P}_{12}\}$ and \emph{uretprobes} $\{\mathbb{P}_4, \mathbb{P}_8, \mathbb{P}_{11}, \mathbb{P}_{15}\}$ to \emph{execute\_*} \{timer, subscription, service, client\} functions to get start and end times of callbacks.
	\item We devise a technique to read the source timestamp (\emph{srcTS}) of a particular data in a topic by probing \emph{rmw\_take\_*} \{int, request, response\} both at entry and exit $\{\mathbb{P}_6, \mathbb{P}_{10}, \mathbb{P}_{13}\}$. \emph{srcTS} is passed by reference to this function and the value is not known when we enter the function since it is determined by calling other low-level DDS functions. Hence, we store the address of \emph{srcTS} at the function entry in a BPF map and read the value from the stored address at the function exit.   
	\item In ROS2, a service can be invoked from multiple client nodes. The response for a particular service request is sent to each client node and, hence, we get events for $\mathbb{P}_{12}$, $\mathbb{P}_{13}$, and $\mathbb{P}_{15}$. However, the client callback is dispatched only in the caller node and to distinguish that we probe \emph{take\_type\_ erased\_response} at exit using $\mathbb{P}_{14}$ and read its return value. If it returns 1 then the client callback will be dispatched.
	\item We have noticed a library named \emph{message\_filters} in AVP. It provides APIs to synchronize data arriving at a node in different topics---used for sensor fusion. 
	We attach a probe $\mathbb{P}_{7}$ to identify a subscriber callback used for data synchronization, i.e., the probed function runs every time data is read from the topic and it needs to be synchronized.  
\end{itemize}
  
\subsection{Trace points in the OS scheduler} %We can attach trace points to different OS scheduler events, e.g., \emph{sched\_switch}, \emph{sched\_wakeup}, and \emph{sched\_process\_fork}.
We attach a tracepoint to \emph{sched\_switch}---an OS event that notifies when the scheduler gives a CPU to a new thread.
%Nevertheless, we plan to extend our framework for trace-based debugging and optimization of ROS2 applications where other scheduler events also become crucial.
%A \emph{sched\_switch} event is generated when the scheduler gives a CPU to a new thread.
Hence, from such an event, we get (i)~the CPU where the switch is happening; (ii)~the PID and the scheduling priority of both previous and new threads; and (iii)~the state of the previous thread when the switch occurs.
These events are necessary to measure a callback's execution time, considering that between its start and end (notified by ROS2 events), the thread running it might get preempted or need to wait for data or a signal.
%Also, if a ROS2 node is pinned to a CPU or it has been assigned a real-time priority, we can get such information from \emph{sched\_switch} events.
Also, we get the configured CPU affinity and scheduling priority of ROS2 nodes from these events. 
We note that if we record all \emph{sched\_switch} events, the memory footprint of the trace data will be too high, e.g., hundred megabytes of data per second.
%An important challenge here is that the memory footprint of trace data will be too high if we record all \emph{sched\_switch} events, e.g., it can generate a hundred megabytes of data in a second.
We reduce the memory footprint by an order of three or more by filtering these events based on the PIDs of ROS2 nodes. 
%Hence, we filter these events by PIDs of ROS2 nodes 
We get these PIDs using the probe $\mathbb{P}_1$ that are then shared with \emph{sched\_switch} event handler using BPF maps.
%This reduces the memory footprint by an order of three or more.      

\setlength{\textfloatsep}{0pt}

\begin{algorithm}[t]
	\caption{Extract callback attributes for a ROS2 node} 
	\label{alg:callbackattrr}
	\footnotesize
	\begin{algorithmic}[1]
		\Require {\emph{PID} of the ROS2 node, \emph{ROSEvents}, \emph{SchedEvents}}
		%\Ensure \emph{CBlist}
		\State \emph{CBlist} $=[\hspace{1mm}]$ 
		\For {\emph{event} in \emph{ROSEvents}(\emph{PID})\emph{.\textbf{SortByTime}}()}
		\If {\emph{event.type} is \emph{CB\_start} ($\mathbb{P}_2$/$\mathbb{P}_5$/$\mathbb{P}_9$/$\mathbb{P}_{12}$)}
		\State record \emph{CB}\emph{.type} \Comment{$\mathbb{P}_2$, $\mathbb{P}_5$, $\mathbb{P}_9$, $\mathbb{P}_{12}$ denotes different CB type}
		\State \emph{CB.start} = \emph{event.time}
		\ElsIf {\emph{event.type} is \emph{timer\_call} ($\mathbb{P}_3$) \textbf{and} \emph{CB.start} $\neq \emptyset$}
		\State \emph{CB.ID} = \emph{event.ID}
		\ElsIf{\emph{event.type} is \emph{take} ($\mathbb{P}_6$/$\mathbb{P}_{10}$/$\mathbb{P}_{13}$) \textbf{and} \emph{CB.start} $\neq \emptyset$}    
		\State \emph{CB.ID} = \emph{event.ID}		
		\If {\emph{event.type} is \emph{take\_response} ($\mathbb{P}_{13}$)}
		\State \emph{CB.intopic} = \emph{\textbf{cat}}(\emph{event.topic}, \emph{CB.ID})
		\ElsIf {\emph{event.type} is \emph{take\_request} ($\mathbb{P}_{10}$)} 
		\State \emph{CB.intopic} = \emph{\textbf{cat}}$\big($\emph{event.topic}, \emph{\textbf{FindCaller}}(\emph{event}, \emph{ROSEvents})$\big)$
		\Else
		\State \emph{CB.intopic} = \emph{event.topic}
		\EndIf
		\ElsIf {\emph{event.type} is \emph{dds\_write} ($\mathbb{P}_{16}$) \textbf{and} \emph{CB.start} $\neq \emptyset$}
		\If {\emph{event.topic} is a \emph{service\_request}} 
		\State \emph{top\_out} = \emph{\textbf{cat}}$\big($\emph{event.topic}, \emph{CB.ID}$\big)$
		\ElsIf {\emph{event.topic} is a \emph{service\_response}}
		\State \emph{top\_out} = \emph{\textbf{cat}}$\big($\emph{event.topic}, \emph{\textbf{FindClient}}(\emph{event}, \emph{ROSEvents})$\big)$
		\Else
		\State \emph{top\_out} =  \emph{event.topic}
		\EndIf
		\State append \emph{top\_out} to \emph{CB.outtopic}
		\ElsIf {\emph{event} shows  \emph{will\_not\_dispatch\_client} ($\mathbb{P}_{14}$)}
		\State \emph{CB.*} $= \emptyset$
		\ElsIf {\emph{event} shows \emph{sync\_subscribe} ($\mathbb{P}_{7}$) \textbf{and} \emph{CB.start} $\neq \emptyset$}
		\State \emph{CB.isSyncSubscriber} = TRUE
		\ElsIf {\emph{event.type} is \emph{CB\_end} ($\mathbb{P}_4$/$\mathbb{P}_8$/$\mathbb{P}_{11}$/$\mathbb{P}_{15}$) \textbf{and} \emph{CB.start} $\neq \emptyset$}
		\State \emph{CB.end} $=$ \emph{event.time}
		%\State \emph{CB.RT} = \emph{CB.end} $-$ \emph{CB.start}
		\State \emph{CB.ET} = \emph{\textbf{GetExecTime}}$\big($\emph{CB.start}, \emph{CB.end}, \emph{PID}, \emph{SchedEvents}$\big)$
		%\State \emph{CB.WT} = \emph{\textbf{GetWaitTime}}$\big($\emph{CB.start}, \emph{CB.end}, \emph{PID}, \emph{SchedEvents}$\big)$		
		\State \emph{CBlist\textbf{.AddToCallback}}(\emph{CB})
		\State \emph{CB.*} $=\emptyset$ 
		\EndIf
		\EndFor
		\State \Return{CBlist}
	\end{algorithmic} 
	%\vspace{-3mm}
\end{algorithm}

\begin{comment}
\begin{algorithm}[t]
	\caption{Compute waiting time---\emph{\textbf{GetWaitTime}}($\ldots$)}
	\label{alg:waittime}
	\footnotesize
	\begin{algorithmic}[1]
		\Require \emph{start}, \emph{end}, \emph{PID}, \emph{SchedEvents}
		\Ensure \emph{WaitTime}
		\State \emph{WaitTime} = 0
		\For {\emph{event} in \emph{SchedEvents}\emph{\textbf{.SortByTime}}}
		\If {\emph{start} $<$ \emph{event.time} $<$ \emph{end}}
		\If {\emph{event.type} is \emph{sched\_wakeup} \textbf{and} \emph{event.pid} == \emph{PID}}
		\State \emph{state} = ``waiting''
		\State \emph{wait\_start} = \emph{event.time}   
		\ElsIf {\emph{event.type} is \emph{sched\_switch} \textbf{and} \emph{event.prev\_pid} == \emph{PID}}
		\If {\emph{event.prev\_state} == 0}
		\State \emph{wait\_start} = \emph{event.time}
		\State \emph{state} = ``waiting''
		\EndIf
		\ElsIf {\emph{event.type} is \emph{sched\_switch} \textbf{and} \emph{event.next\_pid} == \emph{PID}}
		\If {\emph{state} == ``waiting''}
		\State \emph{WaitTime} = \emph{WaitTime} + $\big($\emph{event.time} - \emph{wait\_start}$\big)$
		\EndIf
		\State \emph{state} = ``running''
		\EndIf
		\ElsIf {\emph{event.time} $>$ \emph{end}}
		\State \Return {\emph{WaitTime}}
		\EndIf
		\EndFor
	\end{algorithmic} 
\end{algorithm}

\end{comment}

%\begin{algorithm}
%	\caption{Find event chains}
%\end{algorithm}

%\subsection{Timing model synthesis for ROS2-based applications}
\section{Timing Model Synthesis}\label{sec:modelsyn}
%In a ROS2 node, the single-threaded executor runs a callback from its start until its end before looking into the ready queue of callbacks again. 
As mentioned in Sec.~\ref{sec:semantics}, we use a single-threaded ROS2 executor.
Hence, all events between a \emph{callback\_start} event and the next \emph{callback\_end} event related to a particular ROS2 node---identified by its PID---provide information about one execution of a specific callback (CB), also referred to as a CB instance hereafter. 
%Note that \emph{*} gives the type of callback (see Table~\ref{tbl:events}).
Exploiting the above, we propose Alg.~\ref{alg:callbackattrr} to identify all CBs in a ROS2 node and extract their architectural and timing attributes.
The algorithm requires: (i)~the node's \emph{PID}, (ii)~all ROS2 events, \emph{ROSEvents}, and (iii)~all \emph{sched\_switch} events, \emph{SchedEvents}.
It traverses all ROS2 events pertaining to the given \emph{PID} in the chronological order (lines 2 -- 33) and collects information as follows:
%It collects information from different types of events as follows:
\begin{itemize}[leftmargin=*]
	\item It gets the CB's type and its start time from a \emph{CB\_start} event (lines 3 -- 5). The type is identified from the probe name, e.g., a $\mathbb{P}_2$'s event notifies that a \emph{timer} CB starts.
	\item It gets the CB's ID (lines 6 -- 9) (i)~from a \emph{timer\_call} event for a timer CB and (ii)~from a \emph{take} event for other CBs. 
	%\item For a timer callback, it encounters a \emph{timer\_call} event that shows the callback ID (lines 6 and 7).
	%\item For other callbacks, it encounters a \emph{take} event and the callback ID is obtained from it (lines 8 and 9). 
	%input topic name is obtained from the event's fourth parameter 
	\item The subscribed topic (if any) is obtained from a \emph{take} event (lines 10 -- 15). 
	\begin{itemize}
	\item For a client CB, the service response is read from the topic. To distinguish between responses to different clients, the CB's ID is con\emph{\textbf{cat}}enated to the topic name (lines~10~--~11).
	\item For a service CB, the request is read from the topic. To distinguish between requests from different callers, the caller's ID is con\emph{\textbf{cat}}enated to the topic name (lines~12~--~13). To identify the caller, the algorithm first finds and saves the PID of the \emph{dds\_write} event with the same topic and source timestamp as \emph{take}. Then, it finds the \emph{timer\_call} or \emph{take} event with the same PID, chronologically preceding \emph{dds\_write} and after the last \emph{CB\_start}. This event then provides the ID of the caller CB.     
	\end{itemize}
	\item The published topic (if any) is obtained from a \emph{dds\_write} event (lines 16 -- 23).
	%If the callback publishes on a topic, it encounters a \emph{dds\_write} event and it gets the output topic name from it (lines 16 -- 23). 
	\begin{itemize}
		\item If the topic is used to request a service, it con\emph{\textbf{cat}}enates the CB's ID to the topic name (lines~17~--~18). When there are multiple callers of the same service, this helps to uniquely relate the topic to the caller CB.
		\item If the topic is for service response, it finds the client CB and con\emph{\textbf{cat}}enates its ID to the topic name (lines 19 -- 20). 
		For this, the algorithm finds the \emph{take\_response} event with the same topic and source timestamp as \emph{dds\_write}. There can be $n_{cl}$ such events when there are $n_{cl}$ clients of the same service. For each such event, the algorithm finds the chronologically next \emph{take\_type\_erased\_response} event with the same PID to evaluate if the client CB will be dispatched. Now, in case it is evaluated to be true, the client CB ID is obtained from the \emph{take\_response} event. 
	\end{itemize} 
	\item If a client CB is not dispatched (line 24)---as shown by the \emph{take\_type\_erased\_response} event---the information about the current instance is not stored (line 25).
	%A \emph{take\_type\_erased\_response} event shows if a client CB function will be dispatched (line 24). If it is not dispatched, the information about the current CB instance is not stored (line 25).
	\item A subscriber CB is marked appropriately if it is used for data synchronization---as identified by a \emph{sync\_subscribe} event (lines 26 and 27).
	%It encounters a \emph{sync\_subscribe} event for a subscriber callback that is part of data synchronization. It marks the callback instance accordingly (lines 26 and 27).
	\item The end time of a CB instance is obtained from a \emph{CB\_end} event (lines 28 -- 29). Also, its \emph{execution time} is calculated using Alg.~\ref{alg:exectime} (line 30).
	%Once it gets the end time, it computes the \emph{execution time} (line 30) of the callback instance using Algorithm~\ref{alg:exectime}.  
\end{itemize}
%With the obtained timing attributes, input and output topics, ID, and type, 
At the end of a CB instance, the stored information is added to \emph{CBlist}, a list of callbacks (line 32). 
A new entry is added only if none of the existing entries identifies the same CB as the current instance otherwise the execution time is recorded and the published topic list is updated (if a new topic is encountered) in the matching entry.
For all CBs except a service, the ID is used for matching, while for a service, both the ID and subscribed topic are matched.
%As mentioned earlier, the same service can be invoked by multiple callers and the input topic of a service callback is modified to distinguish between different callers. 
Note that, for a service CB in \emph{CBlist}, the updated topic name identifies a service request by a particular caller.
After updating \emph{CBlist}, all information related to the CB instance is deleted (line 32).

Alg.~\ref{alg:callbackattrr} returns the list of CBs and their attributes (line 34). 
Note that the statistical information (including the \emph{worst-case}, the \emph{best-case}, and the \emph{average} measured values) can be obtained using the stored values of execution times of each CB instance. 
Also, for a timer CB, the difference between consecutive start times gives its approximate \emph{period} of invocation.  

\setlength{\textfloatsep}{2pt}
\begin{algorithm}[t]
	\caption{Compute execution time---\emph{\textbf{GetExecTime}}($\ldots$)}
	\label{alg:exectime}
	\footnotesize
	\begin{algorithmic}[1]
		\Require \emph{start}, \emph{end}, \emph{PID}, \emph{SchedEvents}
		%\Ensure \emph{ExecTime}
		\State \emph{ExecTime} = 0
		\State \emph{last\_start} = \emph{start}
		\For {\emph{event} in \emph{SchedEvents}\emph{\textbf{.SortByTime}}}
		\If {\emph{start} $<$ \emph{event.time} $<$ \emph{end}}
		\If {\emph{event.prev\_pid} = \emph{PID}}
		\State \emph{ExecTime} = \emph{ExecTime} + $\big($\emph{event.time} $-$ \emph{last\_start}$\big)$
		\ElsIf {\emph{event.next\_pid} = \emph{PID}}
		\State \emph{last\_start} = \emph{event.time}
		\EndIf
		\ElsIf {\emph{event.time} $>$ \emph{end}}
		\State \emph{ExecTime} = \emph{ExecTime} + $\big($\emph{end} $-$ \emph{last\_start}$\big)$
		\State \Return {\emph{ExecTime}}
		\EndIf
		\EndFor
	\end{algorithmic} 
\end{algorithm}
\setlength{\textfloatsep}{3pt}

\smallskip
\noindent\textbf{Execution time measurement:} Alg.~\ref{alg:exectime} measures the execution time of a CB instance. 
It combines ROS2 and \emph{sched\_switch} events. It requires: (i)~the \emph{start} and the \emph{end} times of the CB instance (obtained from ROS2 events); (ii)~the \emph{PID} of the ROS2 node that identifies the thread $\mathcal{T}$ executing the CB; and (iii)~all \emph{sched\_switch} events, \emph{SchedEvents}.
It finds all execution segments between the start and the end of the CB instance by chronologically traversing the events in \emph{SchedEvents} and sum up their time lengths to compute the execution time (lines 3 -- 11).
%Due to limited space, we do not give a line-by-line explanation of the algorithm and we belief it is self-explanatory
We note that when \emph{CB\_start} event is generated, $\mathcal{T}$ is running so the first execution segment starts with this event, hence, line~2.
Following a similar reason, the last execution segment of the CB instance ends when we get the \emph{CB\_end} event, hence, lines 9 -- 11.
Further, if the previous thread in a \emph{sched\_switch} event is $\mathcal{T}$, it marks the end of an execution segment (lines 5 -- 6). 
Conversely, if the next thread is $\mathcal{T}$ then it marks the start of an execution segment (lines 7 -- 8).

%It starts by initializing \emph{ExecTime} to zero in line~1.
%When \emph{CB\_start} event is generated, the thread with the given \emph{PID} is running so the first execution segment starts with this event.
%Hence, \emph{last\_start} is initialized to the \emph{start} of the callback.
%Now, our algorithm chronologically traverses the events in \emph{SchedEvents} (line~3).
%In lines 4 -- 6, it checks if a \emph{sched\_switch} event---in between the start and the end of the callback instance---shows that the previously running thread has the given \emph{PID} then it increments \emph{ExecTime} by the time difference between (i)~the timestamp of this event (it marks the end of this execution segment) and (ii)~\emph{last\_start} (it marks the start of this execution segment).
%In lines 4 and 7 -- 8, it checks if a \emph{sched\_switch} event---in between the start and the end of the callback instance---shows that the next thread to run has the given \emph{PID} then it updates \emph{last\_start} to the timestamp of this event (it marks the start of a new execution segment).
%In lines 9 -- 10, it checks if the timestamp of an event is later than the \emph{end} of the callback instance then it increments \emph{ExecTime} by the time difference between (i)~\emph{last\_start} (it marks the start of the last execution segment) and (ii)~\emph{end} of the callback instance (it marks the end of the last execution segment). 
%Finally, it returns the execution time of the callback instance in line 11.  

\smallskip
\noindent\textbf{DAG synthesis:} Based on \emph{CBlist}s generated by Alg.~\ref{alg:callbackattrr} for all ROS2 nodes, we can synthesize their timing model as follows:
\begin{itemize}[leftmargin=*]
	\item Each CB in a \emph{CBlist} is a vertex in the DAG. For $n_{cl}$ callers of a service, we have $n_{cl}$ vertices because there are that many entries in \emph{CBlist}. 
	This is necessary because the same service is then part of multiple non-intersecting computation chains which shall be appropriately modeled.
	\item If the subscribed topic of a CB $cb_k$ matches a published topic of another CB $cb_{k'}$, we draw an edge from $cb_{k'}$ to $cb_k$ except when $cb_k$ is part of data synchronization.
	Adding edges to and from a service CB follows the same rule as the request and response topic names are also updated for the caller and the client CBs.   
	%Considering that the request and the response topics of a service are already updated in \emph{CBlist} to distinguish between clients, we do not need to do anything different while adding the edges to and from vertices representing service callbacks.
	%\item We can have a scenario where a callback has an input topic which does not match with any of the output topics of the traced callbacks, i.e., the publisher callback might not be in the scope of tracing. In this case, we draw an incoming edge to the callback while the edge does not have a source vertex. Similarly, if an output topic of a callback does not match with any of the input topics of traced callbacks, we draw an outgoing edge from the callback while the edge does not have a sink vertex. 
	\item We have a divergence in the DAG when (i)~a CB publishes on more than one topic or (ii)~it publishes on a topic subscribed by more than one CB.
	\item If $cb_{k'}$ and $cb_{k''}$ publish on a topic that is subscribed by $cb_{k}$ then we mark an `OR' junction at $cb_k$, i.e., $cb_k$ is triggered when either of $cb_{k'}$ and $cb_{k''}$ publishes data on the topic.%--- and we can mark $cb_k$ accordingly.
	\item Two or more CBs in a ROS2 node can be used for data synchronization \emph{MS}$_\alpha$ as marked in \emph{CBlist}. 
	The output of \emph{MS}$_\alpha$ is computed only when new data has arrived on each of the subscribed topics of the CBs.
	Here, we add a vertex $cb^{\&}_k$---marked as an `AND' junction---with incoming edges from all CBs in \emph{MS}$_\alpha$.
	We draw an outgoing edge from $cb^{\&}_k$ to each subscriber CB of a topic on which the CBs in \emph{MS}$_\alpha$ publish data.
	%A CB in \emph{MS}$_\alpha$ may not show the output topic in the corresponding entry in \emph{CBlist}. 
	Note that when the input data to a CB in \emph{MS}$_\alpha$ never arrives last during the synchronization, no published topic is found in the corresponding entry in \emph{CBlist}.  
\end{itemize}

%\subsection{Processor load measurement}

\begin{comment}
\subsection{Extract attributes of ROS2 event chains}

\subsection{Measure processor load}

\subsection{Identify ROS2 computation chains:}
\begin{itemize}
	\item provide algorithm that should works for different types of callbacks
	\item specifically describe the challenges with services and actions and how we handle them (actions have not been handled before, need to check services) 
\end{itemize}

\subsection{Measuring execution time and response time of a callback}
\begin{itemize}
	\item provide algorithm; emphasize on combining kernel and user space events
\end{itemize}

\subsection{Measuring end-to-end latency of computation chains}
\begin{itemize}
	\item how to use sequence number/source timestamp to identify one execution of a chain
	\item provide algorithm
\end{itemize}

\subsection{Measuring processor load}
\begin{itemize}
	\item average load can be misleading while maximum load can give indication of the true real-time behavior of applications
	\item provide algorithm on how to compute average and maximum processor load (this is not ROS2 specific)
\end{itemize}
\end{comment}

\section{Deployment of Our Proposed Framework}\label{sec:deploy}
\begin{figure}[tbp]
	\centering
	\includegraphics[width=\columnwidth]{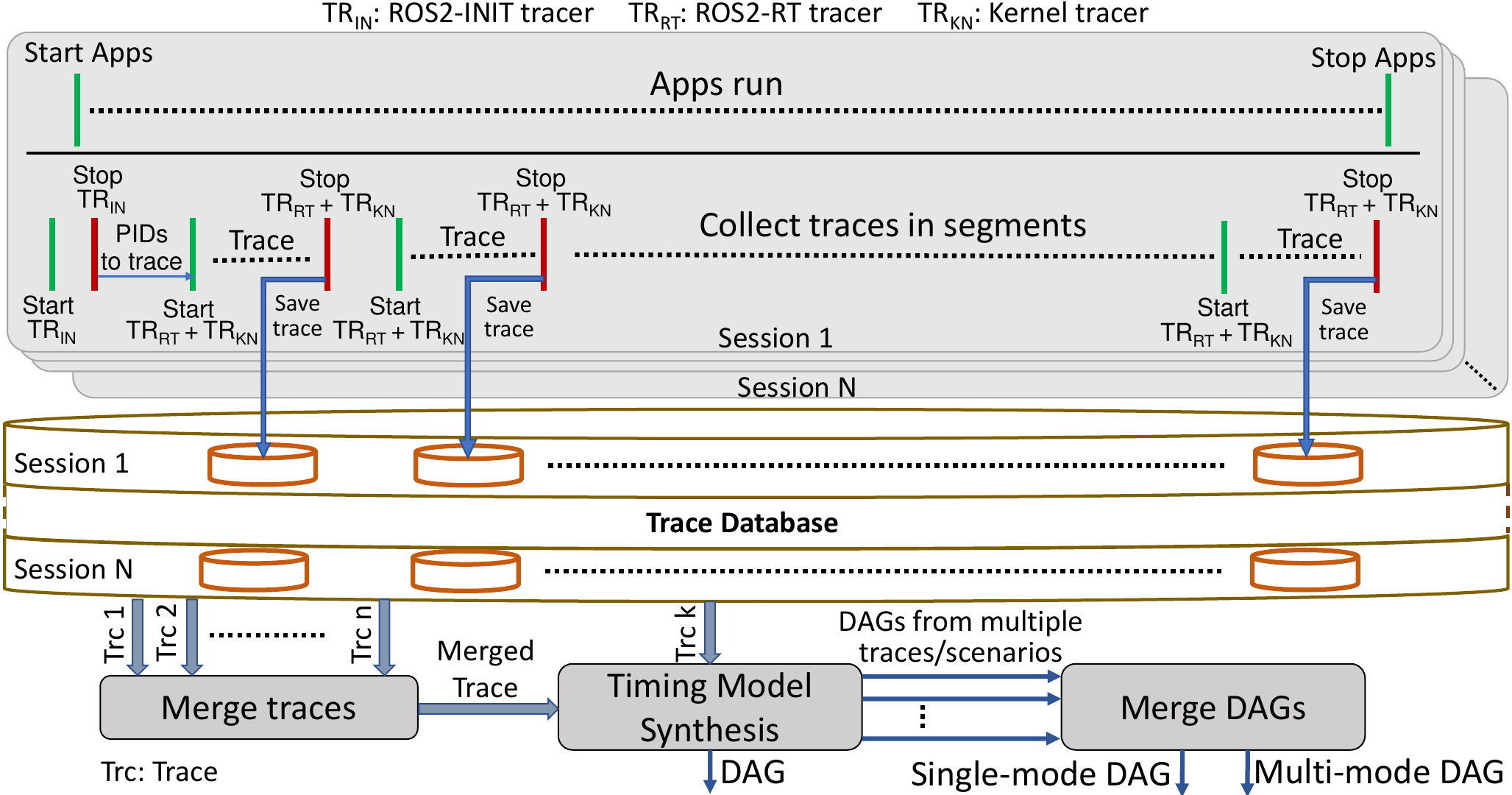}
	\vspace{-4mm}
	\caption {Deployment of tracing and model synthesis framework.}
	\vspace{0mm}
	\label{fig:flow}
\end{figure}

As shown in Fig.~\ref{fig:framework}, we have built three tracers with the eBPF trace points.
(i)~ROS2-INIT tracer ($\text{TR}_{\text{IN}}$) uses $\mathbb{P}_1$ (see Table~\ref{table:ROS2probes}) and traces the initialization of the ROS2 nodes.
(ii)~ROS2-RT tracer ($\text{TR}_{\text{RT}}$) uses $\mathbb{P}_2-\mathbb{P}_{16}$ (see Table~\ref{table:ROS2probes}) and records ROS2 events when the nodes are running. 
(iii)~Kernel tracer ($\text{TR}_{\text{KN}}$) records \emph{sched\_switch} events for ROS2 nodes as identified by $\text{TR}_{\text{IN}}$.
Fig.~\ref{fig:flow} shows how these tracers can be used in practice.
$\text{TR}_{\text{IN}}$ is activated before applications are started and it can be stopped after the initialization.
After $\text{TR}_{\text{IN}}$ identifies the threads of ROS2 nodes, $\text{TR}_{\text{RT}}$ and $\text{TR}_{\text{KN}}$ are activated to record the runtime ROS2 and scheduler events.

In this paper, we suggest synthesizing timing models based on measurements instead of formal worst-case execution time (WCET) analysis considering that the latter does not always scale for industry-strength autonomous applications.
Hence, for accurate modeling, we need to collect large amounts of traces across several runs of the applications and our framework is compatible with this.
Even when one run is long and the trace buffers are limited in size, we can stop $\text{TR}_{\text{RT}}$ and $\text{TR}_{\text{KN}}$, store the traces in a database server, and then restart $\text{TR}_{\text{RT}}$ and $\text{TR}_{\text{KN}}$ with empty buffers.
In the end, we might have a large number of traces in the server collected during multiple tracing sessions, as shown in Fig.~\ref{fig:flow}.

There are several possibilities in which we can process the traces, as depicted in Fig.~\ref{fig:flow}.
(i)~We can merge all traces and run our algorithms for DAG generation on the merged trace.
(ii)~We can generate a DAG for each trace and then they are merged, i.e., the vertices and edges in the final DAG will be a union of all input DAGs while for a callback's execution time, the measured worst-case, best-case, and average values are calculated considering all input DAGs. We take this approach for our experiments in this paper. 
(iii)~We may have a combination of (i) and (ii), e.g., we merge traces collected over one run, generate DAG for each run, and then merge DAGs across multiple runs.
(iv)~Interestingly, it is also possible to generate multi-mode DAGs, e.g., if traces are merged per mode (or scenarios) such as city and highway driving.

\vspace{-3mm}
%\newpage

\begin{figure}[t]
    \centering
    \begin{subfigure}[b]{1\columnwidth}
        \includegraphics[width=1\columnwidth]{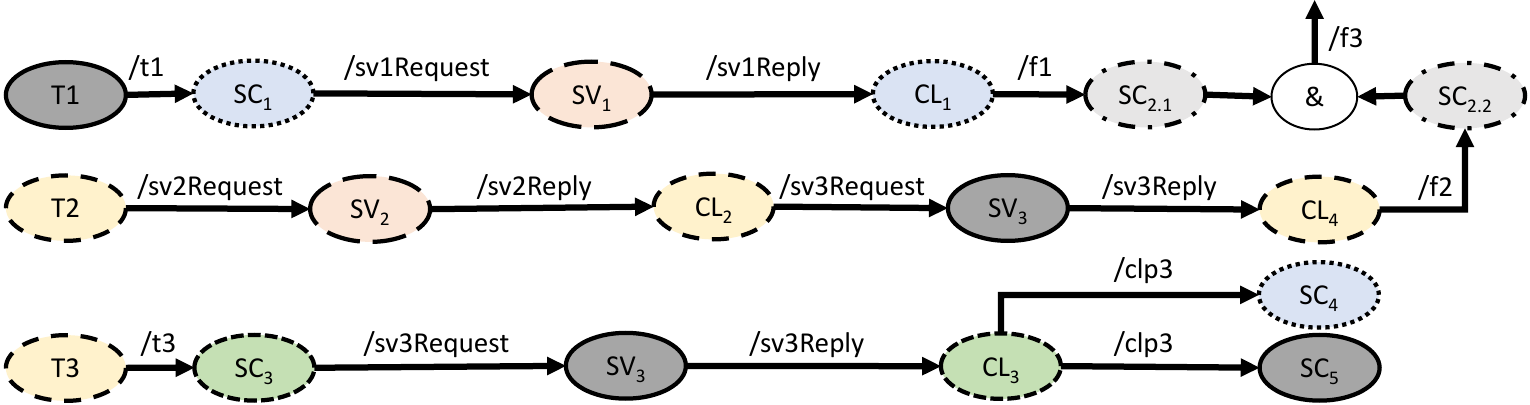}
    	%\vspace{-7mm}
    	\caption {Synthetic application.}
    	\label{fig:synthDAG}
    \end{subfigure}
     %\vspace{2mm}
    \begin{subfigure}[b]{1\columnwidth}
    	\includegraphics[width=1\columnwidth]{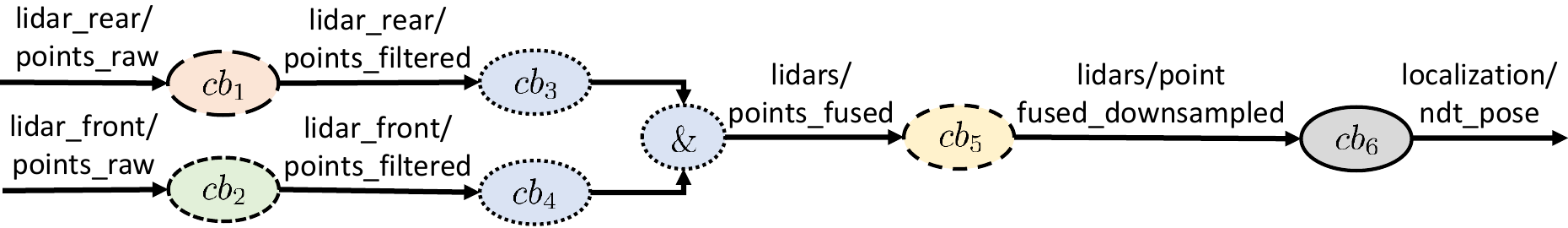}
    	%\vspace{-7mm}
    	\caption {AVP localization.}
    	\label{fig:AVPDAG}
    \end{subfigure}
    \caption{Callbacks and the precedence relations between them.}
\end{figure}

%\begin{figure}[tbp]
%	\centering
%	\includegraphics[width=1\columnwidth]{figures/AVPDAG.pdf}
%	\vspace{-7mm}
%	\caption {Obtained DAG for AVP localization}
%	\label{fig:AVPDAG}
%\end{figure}

\section{A Case Study}
\noindent\textbf{Experimental Setup:}
Our setup comprises ROS2 Foxy, Eclipse Cyclone DDS, and Linux kernel 5.4.1 running on AMD Ryzen 9 3900X CPU with 46~GB RAM.
We run two applications concurrently: (i)~a part of Autoware's \emph{AVP} performing localization~\cite{AVP} and (ii)~a synthetic application (\emph{SYN}).
%The former runs for approximately 80 seconds during which a car from a stationary position drives to a specific parking spot and parks there, thereafter, it comes out of the spot and drives back to the initial position.
%Here, we do not make any modification to the \emph{localization} demo made available by Autoware.  
To run (i), we directly use the \emph{localization demo} provided by Autoware where the application runs for 80 seconds during which a car starts driving from a stationary position, parks, and then drives back to the same initial position. 
Further, we have developed \emph{SYN} using six ROS2 nodes with different combinations of timer, subscriber, service, and client CBs.
We run these applications 50 times, apply our DAG synthesis algorithms on traces collected for each run, and then merge these DAGs together, as explained in Sec.~\ref{sec:deploy}.
%We vary the execution times of CBs in \emph{SYN} across different runs.     

%To evaluate the efficacy of our framework, we developed an experiment on a machine of 12 X CPUs and 46 GB of RAM, running Linux Kernel 5.4.1.
%The setup is composed of two applications running at the same time and bound to the same CPUs.
%The first application is the Autonomous valet parking (AVP) localization demo from Autoware{.}Auto project~\cite{AVP} which is an open-source project that provides a software stack for developing self-driving vehicles built on ROS2.
%The application runs for nearly 80 seconds and demonstrates an autonomous parking functionality during which the car starts from an initial position, drives autonomously to park in a target parking lot and then drives back to the initial position.
%The second application is a synthetic test that has been developed to mimic actual automotive chains with different driving functionalities.
%The synthetic test consists of several nodes that communicate with each other.
%Each node has one or more of the timer callbacks(T$_i$), service callbacks(SV$_i$), client callbacks(CL$_i$), and subscriber callbacks(SC$_i$) that are executing with varying execution time.

\smallskip
\noindent
\textbf{Timing models:} Fig.~\ref{fig:synthDAG} shows the obtained DAG for \emph{SYN}. CBs belonging to the same node in an application are marked with a distinct color and border. Also, the edges are annotated with the topic names. \emph{SYN} covers different scenarios that are correctly identified by our framework.
(i)~It distinguishes between CBs of the same type in a ROS2 node, e.g., T2 and T3 are timer CBs; SC1 and SC4 are subscriber CBs; SV1 and SV2 are service CBs; and CL2 and CL4 are client CBs.
(ii)~It identifies different types of CBs in a node, e.g., T1, SC5, and SV3 are timer, subscriber, and service CBs, respectively.
(iii)~It finds multiple subscribers of a topic, e.g., /clp3 is subscribed by SC4 and SC5.
(iv)~It finds invocation of a service from different CBs, e.g., SV3 is invoked from SC3 and CL2. Also, we have two vertices for SV3, which is how we propose to model such a case. Imagine, if we had only one vertex for SV3 with incoming edges from SC3 and CL2 and outgoing edges to CL3 and CL4, then SC3 $\rightarrow$ SV3 $\rightarrow$ CL4 forms a sub-chain, which is incorrect. 
(v)~It identifies data synchronization, e.g., data from /f1 and /f2 are synchronized and the output is /f3. For timing analysis, the vertex marked as ``\&'' is a task that has a zero execution time and is an ``AND'' junction.

Fig.~\ref{fig:AVPDAG} shows the obtained DAG performing localization in \emph{AVP}.
It consists of 6 CBs in 5 ROS2 nodes. Node names are provided in Table~\ref{table:AVPresults}.
The DAG shows that $cb_1$ and $cb_2$ are in different nodes and \emph{filter LIDAR rear} and \emph{front raw data}, respectively.
The \emph{filtered} data are sent to a \emph{fusion} node where $cb_3$ and $cb_4$ are used for data synchronization, as identified by our framework.
The \emph{fused} data is \emph{downsampled} by $cb_5$ in a \emph{voxel\_grid\_cloud} node.
Finally, $cb_6$ in a \emph{p2d\_ndt\_localizer} node implements the localization algorithm on the \emph{downsampled} data to determine the vehicle's \emph{pose}.
Here, the DAG has only subscriber CBs and an ``\&'' junction in the \emph{fusion} node.
Table~\ref{table:AVPresults} further reports the measured best-case, average, and worst-case execution times (mBCET, mACET, and mWCET, respectively) of each of the 6 CBs. %in \emph{AVP} localization.
They are measured over 50 runs.
Here, the most computationally expensive CB, $cb_2$, has an average processor load of 27\%---the LIDAR data arrives at 10~Hz.
Such measurements are useful even for simple debugging and optimization, e.g., balancing load across processor cores or keeping the load below a certain threshold while determining core bindings of ROS2 nodes. 

\begin{figure}[tbp]
	\centering
	\includegraphics[width=1\columnwidth]{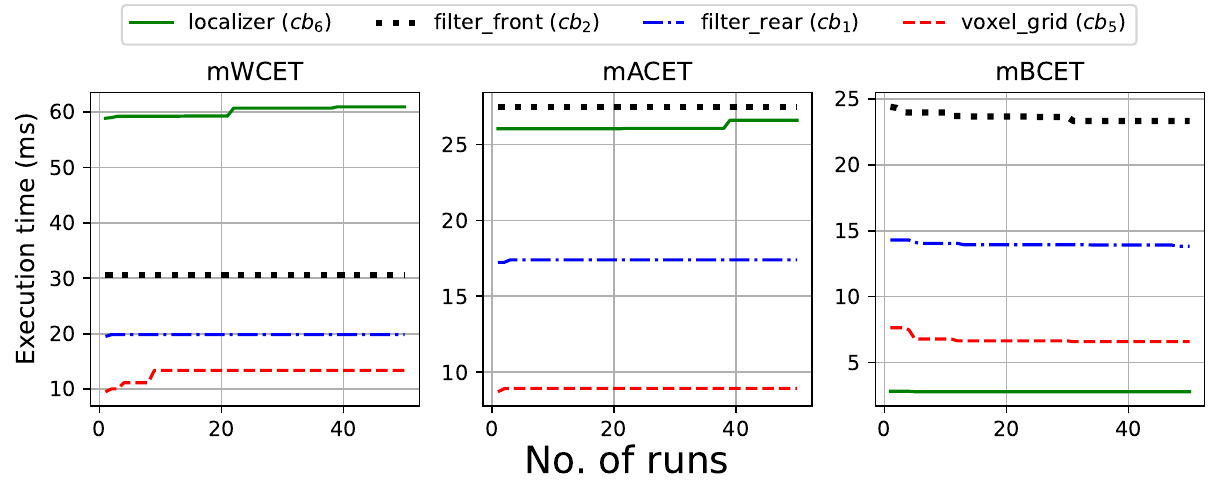}
	\vspace{-6mm}
	\caption {Estimation of timing attributes improve with more traces.}
	\label{fig:evolution}
\end{figure}

For each CB in \emph{SYN}, we have used a constant computational load for a single run. 
By comparing the measured with the designed execution times, we have validated our framework's ability to measure accurately.
We change the load of each CB in \emph{SYN} across runs to evaluate if the execution time profiles of \emph{AVP}'s CB are sensitive to varying interfering loads.
We note that for \emph{SYN}, the measurement results over 50 runs are not important so we do not report them.
%However,  evaluate if the execution time profile of a CB in AVP is sensitive to interfering loads. 
%Hence, we can evaluate if the execution times of a CB in AVP is sensitive to interfering loads.
Further, Fig.~\ref{fig:evolution} shows that mACET and mBCET change negligibly for $cb_2$ with increasing runs, while mWCET increases by 10\% over 23 runs and thereafter remains unchanged.
Such an evolution of mWCET shows that our modeling accuracy improves with more traces.
%We point out that the test case generation to maximize the coverage is orthogonal to this work. 
We point out that, if we assume that test cases can be generated with a high coverage, our framework can support accurate model synthesis using tracing and measurement.

\smallskip
\noindent\textbf{Tracing overheads:} For an experiment where we run \emph{SYN} and \emph{AVP localization} together for 60s, (i)~we generate 9MB of trace data; and %(ii)~the applications use 2.3 CPU cores on average; and (iii)
(ii)~\emph{bpftool} shows that our eBPF probes use 0.008 CPU cores on average, which is 0.3\% of the computational load produced by the applications.

\begin{table}[]
	\renewcommand*{\arraystretch}{1.1}
	\centering
	\caption {Execution times (in ms) of callbacks in AVP localization.}
	\vspace{-2mm}
	\label {table:AVPresults}
	\begin{tabular}{|l|c|c|c|c|}
		\hline  
		\textbf{CB} & \textbf{Node} & \textbf{mBCET} & \textbf{mACET} & \textbf{mWCET}\\
		\hline \hline
		$cb_1$ &  filter\_transform\_vlp16\_rear & 13.82 & 17.1 & 19.82 \\
		\hline
		$cb_2$ &  filter\_transform\_vlp16\_front &  23.31 & 27.07 & 30.5 \\
		\hline
		$cb_3$ & point\_cloud\_fusion & 0.41 & 3.1 & 3.97 \\
		\hline
		$cb_4$ & point\_cloud\_fusion & 0.38 & 0.62 & 3.36 \\
		\hline    
		$cb_5$ & voxel\_grid\_cloud\_node & 6.58 & 8.47 & 13.36\\
		\hline
		$cb_6$ &  p2d\_ndt\_localizer\_node & 2.78 & 25.64 & 60.93 \\ \hline
	\end{tabular}
\end{table}

\begin{comment}
\begin{table}[]
  \centering
  \caption {Extracted Timing Results}
  \label {Table:AVP_results}
  \begin{tabular}{|l|c|c|c|c|}
	\hline  
    Callback & Node & BCET & ACET& WCET\\
    \hline \hline
    CB$_1$ &  filter\_transform\_vlp16\_rear & BCET & ACET & WCET \\
    \hline
    CB$_2$ &  filter\_transform\_vlp16\_front & BCET & ACET & WCET \\
    \hline
    CB$_3$ & point\_cloud\_fusion & BCET & ACET & WCET \\
    \hline
    CB$_4$ & point\_cloud\_fusion & BCET & ACET & WCET \\
    \hline    
    CB$_5$ & voxel\_grid\_cloud\_node & BCET & ACET & WCET\\
    \hline
    CB$_6$ &  p2d\_ndt\_localizer\_node & BCET & ACET & WCET \\ \hline
  \end{tabular}
\end{table}
\end{comment}

\vspace{-3mm}
%\newpage
\section{Concluding Remarks}
\vspace{-0.5mm}
In this paper, we have presented a framework to trace ROS2-based autonomous applications using eBPF. 
It further implements algorithms to process traces and synthesize timing models of applications. 
Also, we have shown how to appropriately model ROS2 services and data synchronization.
We note that our framework can be trivially extended to support other ROS2 versions and DDS implementations, while the proposed concepts can be applied to other software architectures (e.g., AUTOSAR) and operating systems (e.g., QNX).  
In the future, we would like to use the framework for debugging and optimization.
We are logging the source timestamp of data on publisher and subscriber sides using which we can traverse data flow through a computation chain and calculate its end-to-end latency.
We can add a tracepoint to \emph{sched\_wakeup} and compute the waiting time of a callback.
Such measurements will help to debug the timing of a chain and optimize it as per requirements.
Also, along the lines of~\cite{Blass2021RTAS}, we can change the schedule configuration of ROS2 nodes online using eBPF to improve systems' timing performance.

\vspace{-2mm}
\bibliographystyle{IEEEtran}

\bibliography{conference_101719}

\end{document}